\documentclass[aps,prd,twocolumn,floatfix,nofootinbib,superscriptaddress]{revtex4-1}
\usepackage{dcolumn}

\usepackage{amsmath,amsfonts,extarrows,amssymb,mathrsfs,times,bm}
\usepackage{extarrows}
\usepackage{graphicx}
\usepackage{float}
\usepackage{graphicx,epstopdf}
\usepackage{dcolumn}
\usepackage{exscale}
\usepackage{relsize}
\usepackage{mathtools}
\usepackage{mhchem}
\usepackage[colorlinks=true,breaklinks=true,linkcolor=blue,citecolor=blue,urlcolor=blue]{hyperref}
\newcommand{\nn}{\nonumber}

\begin{document}

\title{A Novel Method for Calculating Deflection Angle with Finite-Distance Correction}
\author{Zonghai Li}
\email{sche-me@outlook.com}
\affiliation{Newton-Einstein academy, Jialidun University, Pu’er, 676213, China}

\date{\today}
\begin{abstract}

In recent study in Ref.~\cite{Dashi} (arXiv: 2401.12525), we have introduced a method aimed at calculating the weak-field asymptotic deflection angle. This method offers an efficient computational approach that avoids the complexities of integration and cumbersome iterative procedures typically associated with deflection angle calculations. In the present paper, we expand upon this method to encompass finite-distance deflection scenarios, wherein it is postulated that both the distances from the source to the lens and from the observer to the lens are finite. Importantly, this extension naturally encompasses the case of a lens in asymptotically non-flat spacetime. As an illustrative example, we apply this method to compute the gravitational deflection angle of massive particles in Kerr spacetime while accounting for the effects of finite distance. 

\end{abstract}
\maketitle


\section{Introduction}

Currently, there exist various methods for calculating weak-field limit deflection angles. These different approaches offer diverse perspectives and insights when addressing gravitational deflection problems. They do not inherently possess advantages or disadvantages, and the choice among them depends on the researcher's preferences or specific requirements. Nevertheless, these methods vary in terms of computational efficiency.

The traditional integration method~\cite{Congdon-Keeton} is straightforward and widely applicable (both in weak and strong field regimes); however, their computations are relatively complex and inevitably involve the closest approach distance. The geometric approach, introduced by Gibbons and Werner, which employs the Gauss-Bonnet theorem to calculate the deflection angle, offers an elegant and global perspective on the deflection problem by representing the deflection angle as an integral of curvature~\cite{Gibbons-Werner,Werner2012}. Nevertheless, this method entails cumbersome iterative procedures when computing higher-order deflection angles~\cite{CGJ2019,LWJsd}. Although partial alleviation can be achieved in Cartesian coordinates, it does not obviate the need to solve for lower-order particle trajectories~\cite{LiZhouprd2020}.

For the weak-field asymptotic deflection scenario, Li recently introduced a computational method that avoids complex integrals and iterative solutions~\cite{Dashi}. This method involves the introduction of the straight line $u=1/r=\sin\varphi/b$ and defines a function $\Phi(\varphi)$. The deflection angle is directly dependent on the values of this function at the boundaries, i.e., $\delta=\Phi(0)+\Phi(\pi)-\pi$. Furthermore, through this equation, the deflection angle is expressed directly as a function of the impact parameter $b$. The simplicity of this method lies in the ease of calculating the function $\Phi(\varphi)$.

Due to the computational convenience of this method, our aim is to expand its applicability to other scenarios. In this paper, we specifically concentrate on the finite-distance deflection scenario. The study of finite-distance deflection has seen substantial progress through the utilization of the Gibbons-Werner method. Significantly, these advancements have been built upon the groundwork laid in Refs. \cite{ISOA2016, OIA2017}, as demonstrated by an extensive body of research documented in Refs. \cite{OIA2018, OIA2019, OA2019, CGR2019, LiJiaepjc, LiJiaDuan, PantigA, Huang-Jia, Huang-cao01,Gao-Liu}. The finite-distance deflection aligns more closely with our physical world because it assumes that both the particle source and the receiver (or observer) are at finite distances from the lens. Additionally, in the case of asymptotically non-flat spacetimes, such as Schwarzschild-de Sitter spacetime, the asymptotic deflection angle diverges, but finite-distance deflection angles exist. Therefore, finite-distance deflection, as opposed to asymptotic deflection, holds its unique areas of research significance. It can be anticipated that with enhanced observational precision, the importance of finite-distance effects will further increase. Finally, it is worth noting that when studying the deflection of particles in asymptotically non-flat spacetimes using the Gauss-Bonnet theorem, caution is advised, as modifications to the Gibbons-Werner method are necessary in such cases~\cite{TOA, LZO2020, Huang-cao02, HCL02}.

The aim of this paper is to extend the method introduced in Ref.~\cite{Dashi} to the finite-distance deflection scenario.
The structure of this paper is as follows. In Sec.~\ref{Gou-Fang-Pi}, we establish a general method applicable to finite-distance deflection. In doing so, we employ the Jacobi-Randers metric to derive the particle trajectory equations and provide a review of the computation of asymptotic deflection angles. In Sec.~\ref{Fang-Gou-Pi}, we apply the new method to Kerr spacetime. Specifically, we calculate the second-order finite-distance deflection angle for massive particles in this spacetime. Finally, our paper concludes briefly in Sec.~\ref{Fang-Pi-Gou}. Throughout this paper, we use the geometric units.

\section{Method}

\label{Gou-Fang-Pi}

\subsection{Trajectory equation via $(\alpha_{ij},\beta_i)$}

In Boyer-Lindquist coordinates $(t, r, \theta, \phi)$, the metric for the $4D$ stationary spacetime is given by
\begin{align}
	\label{BL-metric}
	d s^2=&g_{t t}(r,\theta) d t^2+2 g_{t \phi}(r,\theta) d t d \phi+g_{rr}(r,\theta) dr^2\nn\\
	&+g_{\theta\theta}(r,\theta) d\theta^2+g_{\phi\phi}(r,\theta) d\phi^2.
\end{align}

The trajectory of a massive particle with mass $m$ and energy $E$ in the stationary spacetime~\eqref{BL-metric} is the geodesic of the following $3D$ Jacobi-Randers metric~\cite{Chanda2019},
\begin{align}
	\label{Jacobi-Randers}
	F(x,dx)=\sqrt{\alpha_{ij}dx^idx^j}+\beta_idx^i,
\end{align}
where
\begin{align}
	\label{Jacobi-Randers-a}
	&\alpha_{ij}=\frac{E^2+m^2{g}_{tt}}{-{g}_{tt}}\left({g}_{ij}-\frac{{g}_{ti}{g}_{tj}}{{g}_{tt}}\delta_{\phi}^i\delta_{\phi}^j\right),\\
	\label{Jacobi-Randers-b}
	&\beta_i=-E\frac{g_{ti}}{g_{tt}}\delta_{\phi}^i.
\end{align}
It is essential to emphasize that the Randers metric with the form \eqref{Jacobi-Randers} is a specific type of Finsler metric. Here, $\alpha_{ij}$ represents the Riemannian metric, and $\beta_i$ is a one-form, satisfying $\sqrt{\alpha^{ij}\beta_i\beta_j}<1$. 

Introducing the variable $u=1/r$, the trajectory equation for particles on the equatorial plane $(\theta=\pi/2)$ can be expressed in terms of $(\alpha_{ij}, \beta_i)$ as follows~\cite{LWJsd}
\begin{align}
	\label{orbitequation}
	\mathcal{Z}(u)=\left(\frac{du}{d\phi}\right)^2=u^4\frac{\alpha_{\phi\phi}\left[\alpha_{\phi\phi}-\left(L-\beta_\phi\right)^2\right]}{\alpha_{rr}\left(L-\beta_\phi\right)^2}.
\end{align}
Here, $L$ represents the particle's conserved angular momentum, and its relationship with the energy $E$ is described by
\begin{align}
	\label{LED}
	L=bvE,\quad E=\frac{m}{\sqrt{1-v^2}},
\end{align}
where $b$ stands for the impact parameter, and $v$ represents the asymptotic velocity of the particle.

\subsection{Asymptotic deflection angle via $\Phi(\varphi)$}
This subsection provides a concise review of the method proposed in Ref.~\cite{Dashi} for computing the weak-field asymptotic deflection angle of particle on the equatorial plane. 

Utilizing a variable substitution, $u=\sin\varphi/b$, we have
\begin{align}
	\label{zhaoguorong}
\frac{du}{d\varphi}=\pm\sqrt{\mathcal{Z}(u)}\frac{d\phi}{d\varphi}=\frac{\cos\varphi}{b},
\end{align}
where $\mathcal{Z}(u)=(\frac{du}{d\phi})^2$ is the trajectory equation.

Next, we introduce the function $\Phi(\varphi)$ as follows
\begin{align}
	\label{amidabaye01}
	\Phi(\varphi)=&-\left[\int \frac{\cos\varphi}{b\sqrt{\mathcal{Z}[u(\varphi)]}}~d\varphi \right]_{C=0},
\end{align}
where $C$ represents the integration constant.

According to Eqs.~\eqref{zhaoguorong}-\eqref{amidabaye01}, we have
\begin{align}
	\label{sunwukong}
	\phi(\varphi)=\begin{cases}
		-\Phi(\varphi) & \text{if } \varphi < \frac{\pi}{2} \\
		\Phi(\varphi) & \text{if } \varphi >  \frac{\pi}{2}
	\end{cases}.
\end{align}
Note that $\phi(\varphi)$ is undefined at $\varphi=\pi/2$. 

Subsequently, the asymptotic (or infinite-distance) deflection angle can be expressed as~\cite{Dashi}
\begin{align}
	\label{DAF}
	\delta=\Phi(0)+\Phi(\pi)-\pi.
\end{align}

\subsection{Finite-distance deflection angle via $\mathcal{W}(\varphi)$}

Now, let's turn our attention to the finite-distance scenario. Consider a particle on the equatorial plane, originating from the particle source $S$, undergoing deflection by the lens, and ultimately arriving at the receiver $R$. The coordinates of the particle source and receiver are denoted as $(r_S, \phi_S)$ and $(r_R, \phi_R)$, respectively. When $r_S$ and $r_R$ tend to infinity, we return to the asymptotic deflection scenario.

The deflection angles, including both finite-distance and infinite-distance cases, can be defined as follows~\cite{OIA2017}
\begin{align}
	\label{IFdea}
	\delta\equiv \Psi_R-\Psi_S+\phi_{R}-\phi_{S}.
\end{align}
Here, $\Psi_S$ and $\Psi_R$ respectively denote the angles between the tangent direction of the particle trajectory and the rays pointing from the lens to the source and receiver.

We initially compute $\Psi_R$ and $\Psi_S$ using the Riemannian metric $\alpha_{ij}$ given in Eq. \eqref{Jacobi-Randers}, with reference to Ref. \cite{OIA2017}. The unit tangent vector of the particle trajectory in the equatorial plane with coordinates $(r,\phi)$ can be expressed as
\begin{align}
	e^i=\frac{d x^i}{dl}=\left(\frac{dr}{dl},\frac{d\phi}{dl}\right),
\end{align}
where $l$ is the arc length with respect to $\alpha_{ij}$.

Choosing the outgoing direction, the unit radial vector in the equatorial plane can be written as
\begin{align}
	\mathcal{R}^i=\left(\frac{1}{\sqrt{\alpha_{rr}}},0\right).
\end{align}
Then, we have
\begin{align}
	\cos\Psi=&\alpha_{ij}e^i \mathcal{R}^j=\sqrt{\alpha_{rr}}\left(\frac{dr}{dl}\right)\nn\\
	=&\frac{\sqrt{\alpha_{rr}}}{\sqrt{\alpha_{rr}+\alpha_{\phi\phi}(\frac{d\phi}{dr})^2}}~,
\end{align}
where we used $dl=\sqrt{\alpha_{rr}dr^2+\alpha_{\phi\phi}d\phi^2}$. This can be rewritten as
\begin{align}
	\label{ang3}
	\sin\Psi=&\frac{\sqrt{\alpha_{\phi\phi}}}{\sqrt{\alpha_{rr}(\frac{dr}{d\phi})^2+\alpha_{\phi\phi}}}\nn\\
	=&\frac{1}{\sqrt{1+\frac{\alpha_{rr}}{\alpha_{\phi\phi}}\left(\frac{dr}{d\phi}\right)^2}}.
\end{align}

From Eq.~\eqref{orbitequation} we have
\begin{align}
	\left(\frac{dr}{d\phi}\right)^2=\frac{\alpha_{\phi\phi}\left[\alpha_{\phi\phi}-\left(L-\beta_\phi\right)^2\right]}{\alpha_{rr}\left(L-\beta_\phi\right)^2}.
\end{align}
Substituting it into Eq.~\eqref{ang3}, we get a simple result
\begin{align}
	\label{ang45}
	\sin\Psi
	=&\frac{L-\beta_\phi}{\sqrt{\alpha_{\phi\phi}}}.
\end{align}
In the above, we used $L-\beta_\phi > 0$. This is valid because for prograde particle rays, $-\beta_\phi$ is positive, and the angular momentum $L$ from Eq.~\eqref{LED} is also positive. The deflection angle of retrograde particle rays can be derived from that of prograde particle rays~\cite{LWJsd}. Hence, it is sufficient for us to consider only prograde particle rays.

 According to Eq.~\eqref{ang45}, we can express $\Psi_S$ and $\Psi_R$ as follows
\begin{align}
	&\Psi_S=\pi-\arcsin{\frac{L-\beta_\phi}{\sqrt{\alpha_{\phi\phi}}}},\nn\\
	&\Psi_R=\arcsin{\frac{L-\beta_\phi}{\sqrt{\alpha_{\phi\phi}}}}.\nn
\end{align}
We introduce a new function $\mathcal{W}(\varphi)$ as follows
\begin{align}
	\label{amidabaye02}
	\mathcal{W}(\varphi)\equiv\Phi(\varphi)+\left[\arcsin{\frac{L-\beta_\phi}{\sqrt{\alpha_{\phi\phi}}}}\right]_{r=b/sin\varphi},
\end{align}
and then we have
\begin{align}
&\Psi_S=\Phi(\varphi_S)-\mathcal{W}(\varphi_S)+\pi,\nn\\ &\Psi_R=\mathcal{W}(\varphi_R)-\Phi(\varphi_R).\nn
\end{align}
Here and below, we always assume that $0\leq\varphi_S<\frac{\pi}{2}<\varphi_R\leq\pi$.
Using Eqs.~\eqref{sunwukong} and \eqref{amidabaye02}, the deflection angle~\eqref{IFdea} can be written as
\begin{align}
	\label{IFdea-zhongji}
	\delta=\mathcal{W}(\varphi_S)+\mathcal{W}(\varphi_R)-\pi,
\end{align}
where $\mathcal{W}(\varphi)$ is given by Eq.~\eqref{amidabaye02}, and it can be expressed in more detail as
\begin{align}
	\label{amidabaye022024}
	\mathcal{W}(\varphi)=&-\left[\int\frac{\cos\varphi}{b\sqrt{\mathcal{Z}[u(\varphi)]}} ~d\varphi \right]_{C=0}\nn\\
	&
	+\left[\arcsin{\frac{L-\beta_\phi}{\sqrt{\alpha_{\phi\phi}}}}\right]_{r=b/sin\varphi}.
\end{align}
Note that $\mathcal{W}(\varphi)$ is undefined at $\varphi=\pi/2$. Fortunately, we are only concerned with the values of $\mathcal{W}(\varphi)$ at $\varphi_S$ and $\varphi_R$, and we do not need to consider the case of $\varphi=\pi/2$.

Because we start from the general definition~\eqref{IFdea}, the deflection angle~\eqref{IFdea-zhongji} has a broad range of applicability within the weak deflection limit. It is relevant not only for infinite-distance deflection (or asymptotic deflection) but also for finite-distance deflection, and it is applicable to both asymptotically flat spacetimes and asymptotically non-flat spacetimes.

For asymptotic deflection angle in asymptotically flat spacetime, let $\varphi_S=0$ and $\varphi_R=\pi$, we have 
\begin{align}
	&\mathcal{W}(0)=\Phi(0),\nn\\
	&\mathcal{W}(\pi)=\Phi(\pi),\nn
\end{align}
the deflection angle~\eqref{IFdea-zhongji} becomes~\eqref{DAF}.

In addition, $\varphi_S$ and $\varphi_R$ can be respectively expressed as $u_S$ and $u_R$ as follows
\begin{align}
	\label{Dashan}
	\varphi_S=&\arcsin(bu_S),\\
	\label{Beida}
	\varphi_R=&\pi-\arcsin(bu_R).
\end{align}

\section{Application to Kerr Spacetime}
\label{Fang-Gou-Pi}

This section utilizes Eq.~\eqref{IFdea-zhongji} to calculate the deflection angle of massive particle in Kerr spacetime with consideration for finite-distance effect. The model has previously been studied in Ref. \cite{LiJiaepjc} using the Gauss-Bonnet theorem. 

\subsection{Kerr spacetime and Jacobi-Randers metric}

The Kerr metric describes the spacetime outside a rotating body with mass $M$ and angular momentum $J=Ma$. In Boyer-Lindquist coordinates $(t, r, \theta, \phi)$, its line element is given by~\cite{Kerr-BL}

\begin{align}
	\label{kerrmetric}
	ds^2=&-\left(1-\frac{2Mr}{\Sigma}\right)dt^2+\frac{\Sigma}{\Delta}dr^2+\Sigma d\theta^2\nn\\
	&+\frac{1}{\Sigma}\left[\left(r^2+a^2\right)^2-\Delta a^2\sin^2\theta\right]\sin^2\theta d\phi^2\nn\\
	&-\frac{4Mar}{\Sigma}\sin^2\theta dt d\phi,
\end{align}

where
\begin{align}
	\Sigma=r^2+a^2\cos^2\theta,\quad\Delta=r^2-2Mr+a^2.\nn
\end{align}

Substituting the metric components of the Kerr spacetime into Eqs.~\eqref{Jacobi-Randers-a} and \eqref{Jacobi-Randers-b}, and considering the equatorial plane $(\theta=\pi/2,~d\theta=0)$, we obtain the following 2D Kerr Jacobi-Randers metric
\begin{align}
	\label{KerrJR}
	F(r,\phi,dr,d\phi)=\sqrt{\alpha_{rr}dr^2+\alpha_{\phi\phi} d\phi^{2}}+\beta_{\phi}d\phi,
\end{align}
where
\begin{align}
	\label{com-KerrJR1}
	\alpha_{rr}=&\left(\frac{E^2r^2}{\Delta-a^2}-m^{2}\right)\frac{r^2}{\Delta},\\ 
	\label{com-KerrJR2}
	\alpha_{\phi\phi}=&\left(\frac{E^2r^2}{\Delta-a^2}-m^{2}\right)\frac{r^2\Delta}{\Delta-a^2},\\
	\label{com-KerrJR3}
	\beta_{\phi}=&-\frac{2E Ma r}{\Delta-a^2}.
\end{align}

\subsection{The Calculation of $\mathcal{W}(\varphi)$}

Substituting the Kerr Jacobi-Randers data $(\alpha_{ij},\beta_i)$ given in~\eqref{com-KerrJR1}-\eqref{com-KerrJR3} into Eq.~\eqref{orbitequation}, and using $r=1/u=b/\sin\varphi$, we obtain the orbital equation $\mathcal{Z}[u(\varphi)]$. By substituting it into Eq.~\eqref{amidabaye022024} and employing Eqs.~\eqref{LED},~\eqref{com-KerrJR2}, and~\eqref{com-KerrJR3}, we can compute the function $\mathcal{W}(\varphi)$. Expanding up to the second order in $\mathcal{O}(1/b)$, the result is expressed as follows
\begin{align}
	\label{chuntianjiangxin}
	\mathcal{W}(\varphi)=\sum_{j=0}^2\mathcal{W}_j(\varphi)\sqrt{\sec^2\varphi}\left(\frac{1}{b}\right)^j+\mathcal{O}\left(\frac{1}{b^3}\right),
\end{align}
where
	\begin{align}
		\mathcal{W}_0=&-\varphi \cos \varphi+\sqrt{\cos^2\varphi}\arcsin(\sin\varphi),\nn\\ \mathcal{W}_1=&\frac{M}{v^2}\left(1+v^2\right)\cos^2\varphi,\nn\\
		\mathcal{W}_2=&-\left\{3\left(\frac{1}{4}+\frac{1}{v^2}\right) \varphi\cos\varphi-\frac{\sin\varphi}{v^2}\right.\nn\\
		&\left.\bigg[1+\frac{\sin^2\varphi}{v^2}+\left(2+\frac{3v^2}{4}\right)\cos^2\varphi\bigg] \right\}M^2\nn\\
		&-\frac{2Ma}{v}\cos^2\varphi.\nn
	\end{align}

\subsection{Deflection angle}

Based on Eq.~\eqref{chuntianjiangxin}, we can compute $\mathcal{W}(\varphi_S)$ and $\mathcal{W}(\varphi_R)$ and substitute the results into Eq.~\eqref{IFdea-zhongji} to obtain the finite-distance deflection angle. Here, we use Eqs.~\eqref{Dashan}-\eqref{Beida} to express the deflection angle as a function of $u_S$ and $u_R$, approximating the result up to the second order in $\mathcal{O}(1/b)$ as follows

{\small
\begin{align}
	\label{Kerr-result}
	\delta=&\frac{\left(1+v^2\right)\left(\sqrt{1-b^2u_R^2}+\sqrt{1-b^2u_S^2}\right)}{b v^2}M\nn\\
	&+\frac{3\left(4+v^2\right)\left[\pi-\arcsin(b u_R)-\arcsin(b u_S)\right]}{4b^2v^2}M^2\nn\\
	&+\frac{u_S\left[3v^2\left(4+v^2\right)+b^2\left(4-8v^2-3v^4\right)u_S^2\right]}{4bv^4\sqrt{1-b^2u_S^2}}M^2\nn\\
	&+\frac{u_R\left[3v^2\left(4+v^2\right)+b^2\left(4-8v^2-3v^4\right)u_R^2\right]}{4bv^4\sqrt{1-b^2u_R^2}}M^2\nn\\
	&\pm\frac{2a M  \left(\sqrt{1-b^2u_R^2}+\sqrt{1-b^2u_S^2}\right)}{b^2v}+\mathcal{O}\left(\frac{1}{b^3}\right),
\end{align}}
where, the positive and negative signs correspond to retrograde and prograde particle rays, respectively. This result~\eqref{Kerr-result} is exactly equivalent to the one obtained using the Gauss-Bonnet theorem, as presented in Eq.~(42) of Ref.~\cite{LiJiaepjc}. Nevertheless, in that paper we were subjected to laborious and time-consuming computations.

When $u_R\to 0$ and $u_S\to 0$, the finite-distance deflection angle becomes the asymptotic deflection angle, denoted as $\delta_{\infty}$. The finite-distance correction is defined as the difference between $\delta$ and $\delta_{\infty}$, given by~\cite{OIA2017}
\begin{align}
	\Delta\delta\equiv\delta-\delta_{\infty}.
\end{align}

Since the method proposed in this paper yields results consistent with existing ones and exhibits efficient computation, our objective is accomplished. For an analysis of the deflection angle \eqref{Kerr-result} and its finite-distance effects, please refer to Ref.~\cite{LiJiaepjc}.

\section{Conclusion}
\label{Fang-Pi-Gou}
This paper extends the author's recent work~\cite{Dashi}, which proposes a method for computing the weak-field asymptotic deflection angle, to the finite-distance scenario. This extension is straightforward; it only requires replacing $\Phi(\varphi)$ with $\mathcal{W}(\varphi)$, as can be seen by comparing Eqs.~\eqref{DAF} and~\eqref{IFdea-zhongji}. It is worth emphasizing that for asymptotically non-flat spacetimes, the asymptotic deflection angle may diverge, while finite-distance deflection angles exist. Therefore, the Eq.~\eqref{IFdea-zhongji} presented in this paper has a wider range of applicability.

 Our method is applied to Kerr spacetime, and the result is consistent with existing literature. The advantage of our approach lies in its avoidance of difficult integrals and cumbersome iterative procedures, resulting in highly efficient computations. Since the techniques and methods have been well-established, we will refrain from presenting additional examples. Interested readers can utilize this method to investigate particle deflection in asymptotically non-flat spacetimes, such as Schwarzschild (Kerr)-de Sitter spacetime.

As the Spring Festival approaches, the author is preparing to return home to set off firecrackers, reluctantly setting aside the pen in hand, filled with deep reluctance.



\end{document}